\documentclass[showpacs,showkeys]{revtex4}
\usepackage{amssymb}

\newcommand{\bal}{\begin{align}}
\newcommand{\eal}{\end{align}}
\newcommand{\beq}{\begin{eqnarray}}
\newcommand{\eeq}{\end{eqnarray}}
\newcommand{\nneeq}{\nonumber \end{eqnarray}}

\begin{document}
\title{        21st Century Ergonomic Education \\
               From Little {\bf e} to Big {\bf E} } 
\author{       Constance K. Barsky }
\affiliation{  Learning by Redesign, Granville, OH 43023, USA} 
\author{       Stanis{\l}aw D. G{\l}azek  }          
\affiliation{  Institute of Theoretical Physics,
               Faculty of Physics, 
               University of Warsaw } 
\date{         3 March, 2014 }
\begin{abstract}
Despite intense efforts, contemporary educational systems 
are not enabling individuals to function optimally in modern 
society. The main reason is that reformers are trying to 
improve systems that are not designed to take advantage of 
the centuries of history of the development of today's 
societies. Nor do they recognize the implications of the 
millions of years of history of life on earth in which 
humans are the latest edition of learning organisms. The 
contemporary educational paradigm of ``education for all'' 
is based on a 17th century model of ``printing minds'' for 
passing on static knowledge. This characterizes most of 
K-12 education. In contrast, 21st Century education demands 
a new paradigm, which we call Ergonomic Education. This is 
an education system that is designed to fit the students 
of any age instead of forcing the students to fit the 
education system. It takes into account in a fundamental 
way what students want to learn—the concept ``wanting to 
learn'' refers to the innate ability and desire to learn 
that is characteristic of humans. The Ergonomic Education 
paradigm shifts to education based on coaching students as 
human beings who are hungry for productive learning 
throughout their lives from their very earliest days. 
\end{abstract} 



\maketitle
   
\section{\bf SETTING THE STAGE }

The national education reform movement in the USA
(US), which is used here as an example of a highly
developed country, has not delivered the results
envisioned or hoped for when it was launched by {\it A
Nation at Risk} (National Comission on Excellence
in Education, 1983). Rather than rising levels of
national student achievement to meet the needs of
the 21st century, the US reform movement has
instead yielded a growing burden of
responsibilities being piled upon local educators.
Current attempts focused on improving student
performance have not resulted in significant gains
on standardized tests or on graduation rates,
especially for the students who traditionally
perform poorly.

\vskip.1in

Since {\it A Nation at Risk} there has been an increasing 
crescendo of calls for reform in the US. Yet accompanying 
each decade of reform have been numerous reports continuing 
to document the failures of the nation's schools. There 
has been an endless progression of fads sweeping through 
the US schools, each buoyed by inflated claims of its 
power to reshape education for the better. One novelty 
follows another, typically leaving only broken promises 
and dashed hopes behind. The parade of failed {\it pseudo}-innovations 
leaves educators and their communities pessimistic about the 
value of anything new proposed by state or national 
decision-makers. They are fearful that any new developments 
and initiatives will again be temporary -- here today and gone, 
if not tomorrow, then in a year or two -- eliminated because 
of lack of commitment or lack of resources or both. 

\vskip.1in

Changing priorities, overlapping as well as conflicting 
educational targets of achievement and well intentioned 
but poorly structured local, state, and national programs 
have produced an ineffective, inefficient, and increasingly 
irrelevant education system. As one looks across the current 
educational scene it is possible to identify a continuum of 
initiatives that can be observed in most US schools. This 
continuum is made up of at least three general categories: 
chaos, revitalization, and the beginnings of a transformation. 
Some characteristics of each are summarized below.

\subsection{ Chaos }

The first category of the continuum, ``chaos,'' embraces 
most reforms entering classrooms from ``the outside''. These 
initiatives may be mandated by politicians, regulators, 
education administrators, or well-meaning reformers. For 
example, the initiatives in the US in mathematics, English 
and science for quasi national standards are creating 
dilemmas for school districts. Questions being raised 
include: How do we disseminate the standards? How can 
we prepare enough teachers quickly enough to teach the 
new standards? What sort of pedagogy will be most effective 
and how to implement it? How can teachers prepare their 
students for the tests that are to measure student learning 
of the new standards? How do we evaluate teacher effectiveness 
if we do not use the typical value-added measurements? The 
reforms have not only sparked chaos in local schools but 
have actually been counterproductive. They have left parents, 
local policymakers, and especially teachers leery of investing 
in new ideas because the likelihood of failure is far greater 
than of success. Those who recognize the chaos get discouraged 
and stop fighting it. 

\vskip.1in

Meanwhile, newcomers keep emerging certain that the problems 
of schools are easily solved and they jump in to start a new 
round of unrealistic reforms. No one knows what should replace 
the current poorly functioning system or how to phase in a 
replacement such that genuine improvement takes place despite 
the innumerable barriers. Typically these new reforms do not 
result in sustained student achievement. Panaceas such as more 
subject matter courses in college for teachers, more seat time 
for students, or subject specialists in classrooms do not 
address systemic problems of inadequate salaries, poor working 
conditions, inadequate training in how to teach unmotivated 
children, or negative impacts of life in current schools on 
promising young teachers. As Michael Fullan (2007) states ``It 
has become more obvious that the approaches that have been 
used so far to bring about educational change are not working 
and cannot work'' (p. 299).

\subsection{ Revitalization }

But while external reforms are being inflicted on schools 
and teachers, many educators at the local level are creating 
a second kind of intervention. They are creating their own 
innovations to solve the problems they have identified as 
most pressing in their individual schools and classrooms. 
The innovations are ``revitalization'' reforms and have been 
implemented by the people who created them to meet a personal 
need locally. They may be limited to a single classroom when 
a teacher discovers the most effective way to engage the 
students—techniques that may have been gained through 
participation in workshops, collaborative conversations 
with other teachers, or trial and error in the classroom. 
Local initiatives demonstrate enough success to energize 
and revitalize local attempts to improve education, even 
in the face of reforms imposed by outside authorities. 
However, these reforms typically result in incremental 
changes and are rarely recognized or widespread. Furthermore, 
there is typically little research to demonstrate just what 
makes them successful.

\vskip.1in

There are also some revitalization reforms that are more 
widespread than a single classroom. They may have been the 
result of state or nationally funded reforms that have 
demonstrated success in numbers of schools, even nationally. 
Although these reforms in schools may not always demonstrate 
the integrity of the original design, they have had staying 
power because of the engagement by committed program developers 
and teachers. Examples of such reforms in the US include: 
Success for All (Slavin, 2011, 2014), Reading Recovery 
(Clay, 1993, 2005; Lyons and Pinnell, 2001), Physics by 
Inquiry (McDermott and the Physics Education Group, 1995), 
Cooperative Learning (Johnson, Johnson and Holubec. 1994; 
Johnson and Johnson, 2014), National Writing Project (National 
Writing Project, 2014), and Project Lead the Way (PLTW) (2014). 
However, many of these reforms rarely surpass inclusion in more 
than a 1000 of the hundred thousand schools in the US, and even 
with their limited expansion there is little to ensure integrity 
with their original design. However, such reforms do provide 
teachers and schools the opportunity to institute changes in 
classroom structure and instruction that could lead to more 
student engagement.

\subsection{ Transformation }

As locally effective as school or district revitalization 
reforms may be, they typically are not able to effect state 
or national change. There need to be sustainable, scalable 
interventions that enable improvements in learning for all 
students. Such interventions can be referred to as 
``transformations'' and imply new kinds of structures and 
standards of performance. Exceptional revitalization 
innovations currently found in individual teacher's 
classrooms, in schools led by exceptional principals, and 
in districts led by exceptional superintendents (our use of 
the American nomenclature is not important here) may offer 
promising ideas for transformation. Bold experiments are 
called for ``…that generate new and powerful forces, including, 
for example, teachers energies and commitments unleashed by 
altered working conditions and new collective capacities, 
and students' intellectual labor in collaborating with other 
students to do the work of learning'' (Fullan, 2007, p.299). 
However, these need to be integrated with other innovations 
in order to have widespread impact. With time, educational 
transformation could be accelerated to spread over entire 
countries and internationally through competition among 
innovations. 

\vskip.1in

Although the details of the classification are tentative 
and incomplete, it offers some guidelines for assessing 
programs. It would be desirable to reduce unproductive 
activities of the kind listed under {\bf chaos} and make progress 
towards {\bf transformation}. For example several of the ``given'' 
traditional ways for schooling in the US are only historically 
contingent patterns from the beginnings of the 20th century, 
which currently could be challenged. These patterns include: 
the mandate for twelve years in school, the selection and 
arrangement of the curriculum, and the academic credit system. 
Sheppard (2005) specifically points out that the academic 
credit system became antithetical to the ideas of progressive 
education as it expanded beyond the administrative and logistical 
issues for which it was designed. These and other historically 
contingent patterns that are no longer productive for the 21st 
century should be open to challenge. However, recognizing that 
transformations may take many years, shorter-term (three to 
five years) improvements in education are possible with support 
for revitalization programs across more schools.
	
\subsection{ Towards Harmony }

The current focus for both, national, state or regional funding 
is on expanded rigorous testing to determine the degree of success 
in teaching. Unfortunately, what teachers have long known, what 
gets tested is what gets taught. It does not matter whether the 
material is actually preparing students for careers or higher 
education or a fulfilling life. For example, as noted in a 2012 
Ohio newspaper the emphasis has been on ``test scores above learning, 
ratings above reasoning, and appearances above academics.'' The 
editorial asks ``Is that the best plan we have to develop students 
who are prepared to succeed in a complicated, competitive global 
society?'' (Cincinnati Enquirer, 2012). And paraphrasing W. Edwards 
Deming, ``...quality can't be achieved through inspection and sorting 
out defectives, but rather by improving the process...'' (Lillrank, 2010).

\vskip.1in

Confidence in the ability to actually reform education continues to 
erode. Part of the problem is that our education systems are complex 
and messy. The attempts that have been made to understand those 
isolated pockets of success that are always selected as examples 
of what can be accomplished have never been able to account for 
the inability to bring these to a scale beyond a few thousand 
schools. The names may stay but the students rarely experience 
the success of the original model (Education Trust, 1998, 1999; 
Grissmer and Flanagan, 1998). 

\vskip.1in

Research on what impacts student performance continues to be 
ambiguous. There are significant variations in research design, 
problems with establishing control groups, and lack of clarity 
of how teacher preparation and student achievement and performance 
are defined. And there is little research on the interactions of 
the many variables that impact the delivery of quality teaching 
and learning. There is also a lack of historical research on the 
continuity and change that has impacted education.

\vskip.1in

A new paradigm is needed based on the establishment of principles 
for learning and rules for teaching—a set of explicit principles 
that govern the processes of learning and rules that serve as a 
norm for guiding the process of teaching. These rules and principles 
are not susceptible to arbitrary decisions and are increasingly 
recognized as flowing from and incorporating how the brain develops 
and learns as the product of millions of years of evolution 
(Fields, 2010; Gladwell, 2008; Dawkins, 2004) and how it can 
operate most recently in a free society (Dewey, 1997). Some of 
these principles and rules must support the freedom of all students 
to learn the way they are capable of learning, help students realize 
what this means in terms of personal freedom and individual 
responsibility, and safeguard the freedom to choose, prevent 
abuses, and respond to changing contexts.

\vskip.1in

The new paradigm of Ergonomic Education proposed below would allow 
for many competing ideas to fuel students wanting to learn. However, 
these ideas will only be in harmony as long as there is agreement 
on the principles of learning and rules of productive teaching 
studied, recognized, developed and applied by the teachers. With 
freedom of choice in the hands of the learner and rules of productive 
teaching applied by the teacher, new knowledge, innovation, and 
creativity can be fostered to address dramatically changing social 
and environmental contexts. There are reasons to believe that 
harmony of new system building and functioning will emerge from 
the recognition and adoption of the new paradigm of Ergonomic Education.


\section{ HISTORY CEO }

To most people history consists of dates and records of 
events and individuals -- a war, an election, a coup, the 
birth, life and death of notable personae, and more recently, 
the lives and contexts of the general population. In contrast 
to the emphases on specific dates and events, individuals or 
populations, {\it ongoing processes of change} throughout history, 
have had far more profound impacts on people's daily lives 
than do most isolated historical events. 

\vskip.1in

Historians understand that ``the unique and the specific matter 
as much as the universal; that context and initial conditions 
matter; that the world is more complex than what is assumed 
in variable-controlled laboratory experiments; and that 
predictions are at best problematic'' (Staley, 2010, p. 35). 
However, although messy complexities and idiosyncrasies 
characterize the study of history, it is still possible to 
draw inferences that illuminate useful knowledge (Landes, 
Molkyr and Baumol, 2012, p. 528). 

\vskip.1in

Similarities and commonalities in development can be observed 
in the midst of the differences that separate historical events. 
These similarities and commonalities can aid in establishing 
categories and analogies that provide structure for understanding 
the flow of history. This flow of history encompasses {\it processes 
of change over extended periods of time}. A perspective on the 
more profound of the ongoing processes of change that have had 
and continue to have significant impacts are reflected in the 
phrase: ``History CEO''—History Constrains, Enables and Organizes. 
Significant change processes that span centuries can be identified 
and have implications for the future. Understanding the driving 
forces and trajectories of these change processes may help society 
be better prepared to meet the challenges of the 21st century. 
Each change process of History CEO, ``… in and of itself is 
indeterminate, always contingent on numerous factors and usually 
compatible with movement in diverse directions'' (Eisenstein, 2005, p. 333). 
One notable and exceptional example of such diverse paths is the contrast b
etween the impact of the printing press and printing in China compared 
to the impact in Western Europe, particularly in education, over centuries. 

\vskip.1in

More than 1300 years ago China developed and used printing on a large 
scale, but only for a limited time. Initially printing was oriented 
to religious materials and was encouraged by a woman, a prominent 
supporter who became an empress. Religious beliefs embraced by the 
empress {\it enabled} the spread of printing. However, soon after her death, 
printing declined and the process was ignored for two centuries. 
According to Barrett (2008) ``Sheer misogyny was undoubtedly a factor 
in this, but, dynastic politics, religious rivalries, vested scribal 
interests, elite snobbery, and even a whiff of xenophobia also played 
a part to a greater and lesser degree'' (Introduction and Acknowledgements). 
All these conspired to {\it constrain} the spread of printing, despite the 
initial enabling by the empress.

\vskip.1in

In contrast to the history of printing in China, Elizabeth Eisenstein 
(1980) in her two volume ground-breaking history of printing in early 
Western Europe has identified a systematic, centuries-long, change 
and growth for what she has called ``the print culture''. Significant 
and irreversible changes were brought about in Europe following the 
``invention'' of the printing press by Gutenberg (c 1450). 

\vskip.1in

In her book {\it The Printing Press as an Agent of Change: Communications 
and Cultural Change in Early-Modern  Europe} (Eisenstein, 1983), 
Eisenstein explains the now-forgotten constraints that were faced 
by people who had to make do with the oral and scribal culture that 
existed prior to the introduction of printing. Because books were 
very scarce, most individuals were ``constrained'' from gaining knowledge 
from documented sources of that time. Libraries, which were the 
repository of written materials, were few and literally geographically 
out of the range of most individuals. Furthermore, the scribal copying 
process led to repeated and newly introduced inaccuracies over time. 

\vskip.1in

As printed books became more accessible, they {\it enabled} the flow of 
information and the more rapid organization and duplication of data. 
Multiple copies with identical content were more easily produced, 
eliminating unending hours of labor and concentration required by 
scribal copying. The use of books inspired an increasingly widespread 
variety of activities not possible or even imagined prior to the 
existence of the printing press. However, it took more than a century 
and a half after Gutenberg's first publication of Bibles for a ``print 
culture'' to emerge (Eisenstein, 2005). 

\vskip.1in

The centuries-long build-up of activities associated with books both 
{\it enabled} and {\it organized} an increasingly diverse range of applications. 
Eisenstein's examples of advances in print communications led to 
changes in religious cultures and significant developments in early 
science, especially astronomy. The proliferation of books also resulted 
in a paradox of censorship. Banning certain books {\it constrained} the 
dissemination of ideas (particularly those viewed as threatening to 
religious canon) in some quarters while simultaneously {\it enabling} the 
printing of those same books in other quarters specifically so that 
the controversial ideas could be further examined (Eisenstein, 2005, 
pp. 209-285).

\vskip.1in

Although Eisenstein comments that there was undoubtedly an impact 
on thinking and learning as books became ``silent instructors,'' 
carrying their message farther than any public lecture, she does 
not examine the ``print culture'' that became embedded in education. 
Furthermore most educators today do not recognize the considerable 
significance of the centuries-long impact of the printing press 
and printing on the history of education.

\vskip.1in

In 1657, two centuries after Gutenberg introduced the printing press, 
John Amos Comenius, a Czech Moravian teacher, educator, and writer 
completed his {\it Didactica Magna} ({\it The Great Didactic}, 1657, originally 
published in Latin in Amsterdam) ``setting forth the whole art of 
teaching all things to all men'' (Comenius,1992, Cover and Title 
Page).  His book advocated universal education for both boys and 
girls, teaching in the vernacular, the use of textbooks, the 
development of graded schools, pacing of instruction, and specific 
roles for universities. 

\vskip.1in

In Chapter XXXII Comenius (1992) introduces the ``universal and most 
perfect order of instruction'' (pp. 287-294). After a brief introduction 
to the process of printing, Section 5 of the chapter provides analogies 
of teaching to the printing press where ``we might adapt the term 
'typography' and call the new method of teaching `didachography' '' 
(p. 289). He goes on further to describe the following:

\begin{center}

\parbox{14cm}{
``Pursuing this analogy to the art of printing, we will show, by a 
more detailed comparison, the true nature of this new method of ours, 
since it will be made evident that knowledge can be impressed on the 
mind, in the same way that it's concrete form can
be printed on paper'' (p. 289).}

\vskip.1in

\parbox{14cm}{
``Instead of paper we have pupils whose minds have to be impressed 
with the symbols of knowledge. Instead of type we have the class 
books and the rest of the apparatus devised to facilitate the 
operation of teaching. The ink is replaced by the voice of the 
master, since this is what conveys information from the books to 
the mind of the listener; while the press is school discipline, 
which keeps the pupils up to their work and
compels them to learn'' (p. 289).}

\end{center}

\noindent Comenius continues his analogy of comparing students to paper that 
is properly prepared and pressed using ink:

\begin{center}

\parbox{14cm}{
``Similarly, the teacher, after he has explained a construction 
and has shown by examples how easily it can be imitated, asks 
individual pupils to reproduce what he has said and thus show 
that they are not merely learners, but actually possessors of 
knowledge'' (p. 293).}

\end{center}

Section 26 includes discussion of end of year examinations 
which were to be graded by inspectors. These examinations were 
necessary to test the students' knowledge ensuring that the 
``subjects had been properly learned'' (Comenius, 1992, p. 293). 
In 1658, Comenius published {\it Orbis Pictus} or {\it Orbis Sensualium 
Pictus} ({\it The Visible World in Pictures}) (Comenius, 1887). Each 
page has an image illustrating something having to do with the 
natural world such as botany, biology, zoology, religion, human 
activities. Accompanying each image are descriptors written in 
the vernacular (German) and Latin. {\it Orbis Pictus} became a popular 
children's textbook and the model for classroom instruction and 
was translated into many Western European languages by the 1700's. 

\vskip.1in
 
By the mid-19th century Comenius' model of education and instruction 
had been adopted by most of Europe and the US. It provided a uniform 
structure that could address the increasing need of educating more 
and more children with fewer teachers per pupil, providing common 
materials and common curricula across a nation's cultures. It also 
met the workplace needs of the technological and manufacturing 
expansion of the industrial revolution.

\vskip.1in

In 1892, on the three-hundredth anniversary of Comenius' birth, 
there were many celebrations of his work in the US. In the {\it National 
Education Association Proceedings} for 1893 there were several essays 
noting Comenius' impact:

\begin{center}

\parbox{14cm}{
``We have found in Comenius the source and the forecasting of 
much that inspires and directs our education now….What is 
commonplace today was genius three hundred years
ago'' (Hark, 1893, p. 723). }

\vskip.1in

\parbox{14cm}{
``The place of Comenius in the history of education, therefore, 
is one of commanding importance. He introduces the whole modern 
movement in the field of elementary and secondary education. 
His relation to our present teaching is similar to that held 
by Copernicus and Newton toward modern science, and Bacon and 
Descartes toward modern philosophy'' (Butler,
1893, p. 728).}

\end{center}

As the histories of the adoption of Comenius' vision for schooling 
systems globally are examined, commonalities are expected that 
illustrate when history constrained, enabled or organized these 
systems. 

\vskip.1in

Today, 121 years since the celebrations in 1892, most education 
systems and modes of instruction are still embedded in the print 
culture of 1657. The still widely used practices of relying on 
textbooks for conveying knowledge and then assessing learning 
with regurgitation of information via multiple assessments that 
are not informative is outdated. The technology of the internet 
has far outpaced the printed textbook as a means for sharing 
information and gaining static knowledge. And measures of 
performance are inadequate and require development. We believe 
the new paradigm of Ergonomic Education will refocus efforts 
on more productive teaching and learning by taking into account 
in a fundamental way the reasons humans want to learn and how 
they benefit from learning. The section, {\it Why {\bf e} to {\bf E}}, expands 
on the importance of changing the current educational paradigm 
to meet the demands of today and be prepared to adapt to the 
demands of the future.

\vskip.1in

The focus on the print culture illustrates both a paradigm 
shift that occurred from the scribal culture to the print 
culture and the persistence of an education culture that has 
been immune from widespread significant change. In contrast, 
science, technology, and business are conducted very differently 
today compared to the 1600's. What can be learned from the 
constraining, enabling, and organizing principles that have 
impacted their histories? How might these examples contribute 
to understanding the challenges of education reform?

\vskip.1in

For example George E. Smith (2009), a historian of science 
and a philosopher of science, has placed on-line, 
http://www.stanford.edu/dept/cisst/visitors.html, ``Testing 
Newtonian gravity then and now.'' The material reviews the 
historical process of over {\it three centuries} of testing and 
simultaneous refinement of calculations in Newtonian gravity 
as it applies to the solar system. He documents the increasing 
variety of sources of gravitational potentials that have been 
included, from additional objects discovered in the solar system 
to later Einstein's corrections required by special and general 
relativity. But attention to such long-running historical 
processes as the emergence of the print culture or the testing 
of Newtonian gravity still seems to be rare.

\vskip.1in

Thomas Hughes (1983), a historian of technology writes in {\it Networks 
of Power} about the expansion of the electric power system from 
small intercity lighting systems of the late 19th century to the 
regional networks of the 1930's. He identifies phases of (1) 
inventor-entrepreneur, (2) technology transfer, (3) system growth 
and reverse salients, (4) momentum and contingencies, (5) evolving 
systems and planned new ones. These processes mirror the processes 
expressed by our History CEO. Hughes framework enables him to 
compare the development of the electric power systems in the 
United States, Germany, and England. The comparisons along with 
their contextual peculiarities illustrate the differences in 
resources, organizational structures, political climates, 
economic practices, cultural preferences between regions as 
well as nations. The emphasis is on the multifaceted complexity 
of systems and what impacts changes over time. 

\vskip.1in

Landes, Molkyr and Baumol (2012) edited {\it The Invention of 
Enterprise: Entrepreneurship from Ancient Mesopotamia to 
Modern Times}. In contrast to many current books focusing on 
what could be called ``how to be an entrepreneur,'' Landes, 
Molkyr and Baumol wanted to study entrepreneurship and its 
relationship to economic growth from historical accounts. 
In 17 chapters this book describes the socio-technological 
culture in specific periods. Despite variations due to 
context, commonalities in the development of entrepreneurship 
across the centuries could be identified and provide useful 
knowledge that the editors believe have immediate applications 
(pp. 527-528).

\vskip.1in

Peter F. Drucker (1985) also provides a historical 
account of entrepreneurship in his book {\it Innovation and 
Entrepreneurship}. From an examination of both modern and 
historical inventions, including inventions of new organizational 
system structures, Drucker identifies seven sources of innovation 
(1) the unexpected; (2) incongruities; (3) process need; 
(4) industry and market structures; (5) demographics; (6) 
changes in perception; (7) new knowledge. He has identified 
patterns that again have repeated themselves over time in a 
variety of contexts. 

\vskip.1in

Rather than providing a historical narrative portraying 
chaotic and unpredictable interactions of many different 
individuals and events, always in the context of a particular 
historical setting, the above texts search for patterns that 
might be connected through space and time. They are particularly 
interested in patterns that will permit a better understanding 
of the evolution of historical events that are reflected in 
today's environment. These patterns may constrain, enable, 
and/or organize cultures, technologies, political entities, 
etc. over centuries. History CEO attempts distill these 
patterns in a way to help understand the predicaments of 
the current education system. Gaddis (2002) recognizes the 
importance of patterns in history and of ``fitting it all 
together'' (pp.48-49). He does not explicitly address long-running 
processes of change that could exist in the historical record. 
However, he introduces many metaphors comparing how one does 
history to how one does science—particularly the historical 
sciences of astronomy, geology, and paleontology -- all of which 
deal with very long running processes of change. 

\vskip.1in

Significant change processes spanning centuries and multiple 
disciplines can be identified. History CEO offers a vocabulary 
for categorizing these processes into a coherent framework. 
The framework is context dependent, has independent and 
interdependent variables, is open to positive and negative 
(Constraining, Enabling, Organizing) feedback loops, and can 
illustrate the continuity of the change processes. The problems 
of education are both content specific and interdisciplinary, 
they are economic and social, and they have historical roots 
and result from the latest reforms. Comparisons of the processes 
that dominate the success or failure of socio-technological and 
socio-cultural systems thus have the potential of providing 
insights into current education problems and ideas for transformation. 

\vskip.1in

However, time and expertise are needed to collect, analyze, 
and synthesize data. What to do in the meantime? Where to 
begin? There are sufficient though limited examples of education 
successes that could provide the seeds of the new paradigm of 
Ergonomic Education. Engineers, specifically ergonomic/human 
factor engineers interested in education as a foundation of 
the structure and function of society could provide the impetus 
for a new approach for understanding the most pressing problems 
of education. As problem solvers they have been trained to achieve 
specific results in complex environments with available resources. 
Would they be able to analyze the notable education successes and 
failures and subsequently design and produce a transition process 
that would lead from the current education system to the Ergonomic 
Education system of the future?


\section{ A ROLE FOR ENGINEERS }

In order to address a systems approach to educational reform, 
Wilson and Barsky (1999) proposed that the six roles of teacher, 
principal, student, district administrator, consultant, parent 
and the community, involved in educational change as described 
by Fullan (1991), needed to be expanded. Wilson and Barsky 
described a role for engineering research based on an analysis 
of Reading Recovery (Reading Recovery Council of North America, 2014). 
Unlike most reforms, Reading Recovery has been successful in the US 
for over a quarter century. It is an integrated system that has 
maintained its integrity yet has so far not realized its scale-up 
potential. It was proposed that the introduction of engineering 
research would be able to reverse engineer the components contributing 
to Reading Recovery's success and introduce an R\&D mindset that would 
lead to future expansion.

\vskip.1in

Scientists are concerned with understanding how the natural or 
biological worlds work, engineers are concerned with problem-solving. 
Yet both of the disciplines share centuries of development contributing 
to the knowledge and practices of today. In contrast school improvement 
research only began a sustained build-up in the 1960's (Fullan, 1991) 
and school improvement/educational reform appears intellectually 
fragmented and not systemic. 

\vskip.1in

The evidence that an engineering research discipline could significantly 
impact education reform is based on the experience with discipline-based 
research in the sciences, particularly physics (Arons, 1990; McDermott 
and the Physics Education Group, 1996). The recent report by the National 
Research Council (2012), {\it Discipline-Based Education Research: Understanding 
and Improving Learning in Undergraduate Science and Engineering}, emphasized 
the need for this continuing research which complements the work in the 
cognitive sciences by introducing specific attention to the disciplines' 
priorities, knowledges, worldviews and practices. It advocates the 
adoption of evidence-based teaching practices to improve learning 
outcomes for undergraduate science and engineering students. However, 
what constitutes the ``evidence'' depends on the education paradigm. In 
order for the Ergonomic Education paradigm to be realized, there must 
be evidence that it is more successful than the current paradigm in 
creating an environment of productive learning. 

\vskip.1in

The engineering disciplines are poised to accumulate the evidence 
required through their understanding of historical precedents (e.g,. 
aircraft manufacturing: Vincenti, 1990; Newman and Vincenti, 2007), 
knowledge of design (e.g., aircraft evolution from the Wright Brothers 
to the Boeing 777 and Airbus: Vincenti, 1990; Perrow, 1999; Petroski, 
1996, 2006), processes of change (e.g., failures of designs in air 
transport, bridge building, nuclear power plants: Perrow, 1999; 
Petroski, 1996, 2006); the role of constraints (Perrow, 1999;Petroski, 2006) 
and reverse engineering. Ergonomics and Human Factors Engineering in 
particular have ``a unique combination of three fundamental characteristics: 
(1) they take a systems approach (2) they are design driven and (3) they 
focus on two closely related outcomes: performance and well-being'' 
(Dul, Bruder, Buckle, Carayon, Falzon, Marras, Wilson and van der Doeler,  
2012, p. 4). And they bring a ``...holistic, human-centered approach to 
work systems design that considers physical, cognitive, social, 
organization, environmental and other relevant factors.'' 
Karwowski (2005, p. 437). Karwowski goes on to point out that 
organizational ergonomics is concerned with the optimization of 
socio-technical systems, including their organizational structures, 
policies and processes.

\vskip.1in

One can ask if the ergonomics/human factors engineering disciplines 
are willing and interested in accepting the challenge of bringing 
about a transformation in education which could address the concerns 
raised by Dul, Bruder, Buckle, Carayon, Falzon, Marras, Wilson and 
van der Doeler (2012) that ``Human factors/ergonomics (HFE) has great 
potential to contribute to the design of all kinds of systems with 
people (work systems, product/service systems), but faces challenges 
in the readiness of its market and in the supply of high-quality 
applications'' (Abstract). However, if the ergonomic engineering 
system design were applied to change educational systems and showed 
some promise for overcoming the roadblocks inherent to Comenian 
design, ergonomics/human factors would certainly get a firm hold 
in broadly understood contemporary professional culture as a resource 
of great importance.
   

\section{ WHY $\rm \bf e$ TO $\rm \bf E$ }
 
The issue of education in the 21st century was an important 
subject of the first STHESCA conference in Kraków in July 
2011 (STHESCA, 2011). Preparations included a seminar also 
held in Kraków in November 2010 that served the purpose of 
defining a starting point for discussing the subject of 
science, technology, and higher education in contemporary 
society (Glazek, 2011). The proposed starting point obtained 
the working title ``{\bf e and E}.'' Lower-case, or small {\bf e }denotes 
the contemporary world-wide system of education rooted in the 
design by Comenius from the 17th century (Comenius, 1992). 
Upper-case, or big {\bf E} denotes a system on a par with contemporary 
needs.

\vskip.1in

The table below gives examples of features of {\bf e }that, in 
comparison with features of {\bf E}, appear already outdated. By 
the visible contrast, these examples also illustrate that 
improvement of {\bf e }does not automatically lead to creation 
of {\bf E}. An explanation of the entries in the table is provided 
below.
\begin{widetext}
\begin{center}
\begin{table}[ht]
\caption{ \label{tabela} 
Examples illustrating differences between systems
{\bf e}
and {\bf E} (Glazek, 2011).
Explanations for the entries are provided 
below item by item. } 
\begin{tabular}{|c|l|l|}
\hline
~~~nr~~~  & ~~~~~~~~~~~~~~~~~~~~~~~~~~~~{\bf e}           
    & ~~~~~~~~~~~~~~~~~~~~~~~~~~~~~~~~  {\bf E}     \\                                   
\hline 
\hline
1 & ~~~subject matter                       
  & ~~~person                                       \\         
2 & ~~~curriculum                           
  & ~~~context                                      \\             	
3 & ~~~focus on student weaknesses                  
  & ~~~focus on student strengths                   \\     
4 & ~~~separation of values from subject    
  & ~~~natural connection of subject with values~~~ \\ 
5 & ~~~one-size-fits-all testing for grades                   
  & ~~~individualized informative assessment        \\
6 & ~~~passing exams                        
  & ~~~performance                                  \\           
7 & ~~~Comenian system                      
  & ~~~post-Comenian system                         \\
8 & ~~~lack of accountability for learning~~~~~~~~~~~
  & ~~~accountability for learning \`a la RR        \\
9 & ~~~teaching according to age           
   & ~~~individualized teaching, life-long          \\
10 & ~~~no self-correction system           
   & ~~~systematic self-correction system           \\
11 & ~~~compulsion                           
   & ~~~one's will                                  \\             
12 & ~~~out of focus                                  
   & ~~~life-long brain development                 \\
13 & ~~~out of focus                                  
   & ~~~ten thousand hours                          \\        
\hline
\end{tabular}     
\end{table}
\end{center}
\end{widetext}

\begin{enumerate}
\item
 System {\bf e } is focused on teaching subject matter, whereas 
   {\bf E} is focused on educating a person in the context of a subject.
\item
 In {\bf e}, the dominant form of teaching is specified by a 
   curriculum independently of the context of students' 
   lives. In {\bf E}, a context important to students is a natural 
   stimulus for learning important concepts.
\item
 In {\bf e}, students are punished if they do not know, do not 
   understand, or cannot do something, until they fulfill 
   the requirements, even if only superficially. In {\bf E}, 
   students improve upon what they are good at, and this 
   is how they notice new elements and directions worth studying.
\item
 In {\bf e}, teaching subject matter is disconnected from teaching 
   values and building character. Natural sharing of useful 
   information about the world among people in a group in {\bf E} 
   replaces destructive competition (Bok, 1987) and teaches 
   principles of understanding in making decisions and handling 
   resources.
\item
 One-size-fits-all testing for grades in {\bf e }is replaced in {\bf E} 
   by providing feedback regarding individual progress in 
   skill acquisition.
\item
 Testing of short-term memorizing ``to get credit'' in {\bf e }is 
   replaced in {\bf E} by assessment of student performance in practice, 
   akin to how skills of all other members of the system {\bf E} are 
   assessed (Drucker and Maciariello, 2008).
\item
 Comenius designed the process of teaching students in {\bf e }as 
   analogous to printing books in a press, while {\bf E} fulfills 
   contemporary requirements (Drucker, 1993). 
\item
 Reading Recovery (Reading Recovery Council of North America, 2014) 
   has a system for monitoring teachers' work in terms of their 
   students' progress in acquiring skills. This is worth studying 
   as a candidate for use in E; there is no such system in {\bf e }
   (Kenneth G. Wilson, personal communication).
\item
 {\bf e }functions like a production line ordered according to age, while 
   {\bf E} accounts for differences among students, enabling them to develop 
   over the lifespan (Drucker, 1993).
\item
 {\bf e }becomes outdated and fails, having no system of self-correction, 
    while {\bf E} is by definition being created so that it changes in agreement 
    with the needs of its clients (Wilson and Daviss, 1994; Wilson and Barsky, 1998).
\item
 {\bf e }is based on compulsion, and {\bf E} on students' will to learn 
    (Glazek and Sarason, 2006) in agreement with the hypothesis 
    (Glazek, 2008) that processes of learning based on will are 
    those that lead to true learning, associated with changes in 
    structure and functioning of the brain and other body parts.
\item
 In {\bf e}, the human brain is treated in practice as a device ready 
    for one-time programming, while in {\bf E} as an organ that grows 
    and changes throughout the lifespan (Fields, 2010; Dawkins, 2004).
\item
 Ten thousand hours is the amount of time of deliberate practice 
    required for reaching an expert level of performance 
    (Ericsson, 2004; Gladwell, 2008) and a teacher needs this 
    much deliberate practice to become a good teacher in {\bf E}.
\end{enumerate}
The explanation of number 7 in the Table says that a sketch 
of specifications for system {\bf E}, congruent with the direction 
of development of the contemporary world, has already been 
drawn by Drucker (1993). During Drucker's nearly century-long 
life he actively studied the practice of management processes 
involved in the transition of the most-developed countries 
from domination of a manual work force through instruction 
to domination of a workforce characterized by mental work 
based on values, knowledge and skills. Specifications for 
the emerging system {\bf E} along with predictable mechanisms of 
creating, principles of measuring (different from the ones 
applied in e), and methods of improving {\bf E} by new generations 
until {\bf e }is almost completely eliminated (probably still in 
the 21st century), form a list of challenges for ergonomic 
engineers. These will be engineers who out of respect for 
human factors are willing to engage in redesigning education 
for the world to have a future.

\vskip.1in

Suppose that inhabitants of the most-advanced countries cease 
to accept systems of type {\bf e }and learn in them less and less 
effectively, while the systems of type {\bf e }enriched with new 
knowledge and technology continue to be very effective in 
developing countries. A question arises: Is not a change 
from {\bf e }to {\bf E} in the leading countries a necessary condition 
for their continued fulfillment of this role?


\section{ ENGINEERING {\bf BY} KIDS }

It is clear from our outline that the task of changing 
education is not to be completed quickly, even if the 
change is a necessary condition for continuation of 
development of democratic social systems. The cohorts 
of ergonomic engineers will have to engage in such tasks 
over many generations. To educate the required generations 
of capable engineers, one has to involve candidates as 
early as when they are children so that they will later 
have a chance to excel over time in systemic handling of 
extremely complex processes of learning. If people continue 
to apply systems of type {\bf e }to educate engineers, there is 
no chance that the alumni will develop the new system {\bf E}. 
In order to educate engineers who will possess the needed 
values, knowledge and skills one has to switch from 
teaching kids according to the principles of engineering 
{\bf for} kids (look, kids, you ought to do this or that) to 
the principles of engineering {\bf by} kids (look, kids, you 
know the problem you face, design a solution). 

\vskip.1in

An illustration of the paradigm shift from {\bf e }to {\bf E} is 
provided by the following which incorporates principles 
of the scouting movement. There is a situation at a 
scouting camp where one group of scouts is separated 
from another by a river. They conceive of a bridge and 
get it built. Instead of imprinting minds in a classroom 
for engineers, knowledgeable and skillful coaches help 
students design and build a bridge and de facto some of 
them become seriously interested in engineering. In 
contrast, a teacher in a classroom could explain to 
students who sit at their desks how carpenters work 
on a bridge. This example does not need further explanation 
in order to catch the attention of serious human factor 
thinkers whose intention is to make the human learning 
as ergonomic as it can be.

\vskip.1in

There exists a great deal of information about the 
principles and functioning of youth organizations that 
provide members with the values, knowledge and skills 
they need. Poland has a particularly rich tradition in 
this respect because of the long national struggle for 
freedom. Unfortunately, the available sources are 
practically solely available in Polish. For completeness, 
two examples are Janowski (2010) and Kami\'nski (2013). 
In addition, contemporary Polish examples of the role 
that scouting experience can play in producing leaders 
of great merit include Micha{\l} Kulesza, (Wikipedia, 2014) 
who engaged in scouting in his youth and later designed 
the democratic government system for Poland after Solidarity 
took power from communists. An example from the US is 
provided by Frances Hesselbein (Hesselbein, 2011). She 
led the Girl Scouts of the US and helped members of the 
American army leadership understand the principles of 
education needed for soldiers and officers. 

\vskip.1in

One should also note the great need for the education 
of managers in values, knowledge and skills stressed by 
Joseph Maciariello (Maciariello and Linkletter, 2011). 
This concerns not only the management of engineering but 
quite generally high level management in all types of 
organizations that comprise society. Maciariello stressed 
the lack of required education for managers as the key 
reason for recent world crises. The references provided 
in this section substantiate a link between the need for 
Ergonomic Education and the well-being of society on a 
large scale that actually originates in how kids are 
educated, with scouting providing examples. These 
examples are also supported by the dialog between 
Drucker and Albert Shanker, the late president of 
the American Federation of Teachers (Drucker, 1990, pp. 132-138).

\vskip.1in

The ergonomic principle of designing the environment to 
fit the user instead of being content with forcing the 
user to fit the environment provides engineers and 
engineering education with the opportunity to engage 
with the big {\bf E} paradigm. These engineers can become 
leaders of the new paradigm. The ergonomic perspective 
of the human condition is here viewed as capable of 
becoming a basis for the new educational design, provided 
that the engineers: (1) recognize the principles of 
productive learning; (2) adhere to the spirit of 
performance defined by Peter Drucker; (3) incorporate 
Drucker's principles of management in organizing their 
new system; and (4) manage the practice of parental 
support for teacher initiatives in building the new 
system of engineering by kids. Children and teachers, 
with support of parents, learn the principles of 
ergonomic engineering through and in collaboration 
with qualified engineers. They experience engineering 
as a vital element of a rational approach to the human 
condition in a highly developed society. The reason for 
calling this section Engineering {\bf by} Kids, is that the 
students are self-motivated to learn, are carefully 
informed by their coaches about the progress they make, 
and are helped by coaches in discovering that they need 
to learn more.


\section{ CONCLUSIONS }

The paradigm shift from {\bf e }to {\bf E} for education is based 
on the history of science, technology and society 
(e.g., Hughes, 1983; Smith, 2009) and the millions 
of years of biological development supporting the 
current learning abilities of humans 
(e.g., Dawkins, 2004; Fields, 2010). The new paradigm, 
big {\bf E}, brings the current condition of educational 
chaos to a transition and transformation that 
recognizes the human condition of wanting to learn. 

\vskip.1in

Ergonomic Education focuses on student strengths not 
weaknesses, it emphasizes the context of learning and 
not a fixed curriculum, it is not constrained by 
age-graded classes, and it envisions a self-correcting 
system for life-long learning. In Ergonomic Education 
the disciplines are no longer dictated to students but 
are discovered, practiced and developed by each new 
generation of students as they develop their strengths 
both as individuals and as part of a team. It incorporates 
the concept of productive learning (Glazek and Sarason, 2006) 
and Drucker's (Drucker and Maciariello, 2008) principles of 
the spirit of performance and continuity, embodied in the 
pattern of practice in Engineering by Kids. It is a system 
that takes a new approach to education by engaging students 
of any age in building their learning habits through what 
they want to learn and helping them to learn to benefit 
from deliberate practice (Ericsson, 2004) as they enter 
and develop the world of values, knowledge and skills. 

\vskip.1in

The Ergonomic Education paradigm shifts to education based 
on coaching students as human beings who are hungry for 
productive learning throughout their lives from their 
very earliest days. All students, regardless of age, are 
helped to learn in a system formed by competent educators 
operating within a new organizational structure. Ergonomic 
engineers are uniquely qualified for designing and building 
such a functioning system of big {\bf E} and continuing to make 
adjustments to improve the system to meet the individual 
learning needs of students and groups of students as they 
strive to learn new knowledge and develop social skills. 

\vskip.1in

In addition to the direct impact of {\bf e }to {\bf E} on education, 
we propose that this is the path to Drucker's (1993) 
concept of continuity in a global society of organizations 
where continuing economic growth, world-wide stability, 
and democratic leadership can flourish. This is a much 
more desirable situation than exists in societies which 
continue to use the Comenian ``printing minds'' education 
system to either direct or even limit the opportunities 
for its citizens to learn. The paradigm of Ergonomic 
Education, in which the foundations of knowledge, skills 
of management, and engineering must be combined to build 
a post-Comenian educational system, appears to be a 
necessary condition for successful leadership of advanced 
democratic societies in the world. 

\vskip.5in

{\bf Acknowledgement:} This paper is the result of a long 
collaboration between the authors and the late 
Kenneth G. Wilson who died as we were in the midst 
of formalizing the ideas presented and preparing a 
series of papers to elaborate on the proposal for a 
new paradigm for education.

\newpage

{\bf REFERENCES }

\vskip.2in

\begin{description}

\item[Arons, Arnold B. (1990).]
{\it A Guide to Introductory 
Physics Teaching}. Hoboken, NJ: John Wiley \& Sons. 

\item[Barrett, T. H. (2008).] {\it The Woman Who Discovered 
Printing}. Hartford, CT: Yale University Press.

\item[Bok, Derek (1987).] {\it Harvard University. The Presidents 
Report 1986-1987}. Harvard University Archive. 
Boston, MA. Retrieved from: http://nrs.harvard.edu/urn-3:hul.arch:14992.  

\item[Butler, Nicholas Murray (1893).] {\it John Amos Comenius, 
exercises in commemoration of the three hundredth 
anniversary of his birth, 1592-1892. The place of 
Comenius in the history of education}. National Education 
Association Journal of Proceedings and Addresses. 
Session of the Year 1892 held at Saratoga Springs, 
New York, 1893, 723-728. Retrieved from 

http://archive.org/stream/addressesproce1892natiuoft\#page/848/mode/2up. 

\item[Cincinnati Enquirer (2012, August 12).] 
{\it A test of confidence: scandals show need to review Ohio's system of standardized 
tests in schools}. Retrieved from 

http://news.cincinnati.com/article/20120811/EDIT01/308110087/A-test-of-confidence.  

\item[Clay, Marie (1993).] {\it Reading Recovery}. 
Portsmouth, NH: Heinemann Press.

\item[Clay, Marie (2005).] {\it Literacy lessons designed for 
individuals part two: Teaching procedures}. 
Portsmouth, NH: Heinemann Press.

\item[Comenius, John Amos (1992, originally published 1657).] 
{\it The Great Didactic of John Amos Comenius}. 
(M. W. Keatinge, Trans. 1910. M.W Keatinge,(Ed.). 
Montana: Kessinger Publishing Company.

\item[Comenius, John Amos (1887, originally published 1658).] 
{\it Orbis Pictus}. (Charles Hoole, Trans.). Charles W. Bardeen, 
(Ed.). Retrieved from http://www.gutenberg.org/ebooks/28299

\item[Dawkins, Richard (2004).] {\it The Ancestor's Tale, a Pilgrimage 
to the Dawn of Evolution}. Boston, MA: Houghton Mifflin.

\item[Dewey, John (1997, originally published 1938).] {\it Experience 
and Education}. New York, NY: Simon \& Schuster.

\item[Drucker, Peter F. (1985).] {\it Innovation and Entrepreneurship, 
Practice and Principles}. New York, NY: Harper \& Row.

\item[Drucker, Peter F. (1990).] {\it Managing the Non-Profit Organization: 
Principles and Practices}. New York, NY: Harper Business, pp. 132-138.

\item[Drucker, Peter F. (1993).] {\it Post-Capitalist Society}. 
New York, NY: HarperCollins Publishers.

\item[Drucker, Peter F. and Joseph A. Maciariello (2008).] 
{\it Management, Revised Edition}. New York, NY: HarperCollins.

\item[Drucker, Peter F. \& Masatoshi Ito Graduate School of 
Management (2014).] Retrieved from 

www.cgu.edu/pages/281.asp. 

\item[Dul, Jan], Ralph Bruder, Peter Buckle, Pascale Carayon, 
Pierre Falzon, William S. Marras, John R. Wilson, and 
Bas van der Doeler (2012). {\it A strategy for human factors/ergonomics: 
developing the discipline and profession}. Ergonomics. 
Retrieved from http://dx.doi.org/10.1080/00140139.2012.661087. 

\item[Education Trust (1998).] {\it Good teaching matters: how 
well-qualified teachers can close the gap}. Thinking K-16 (3.2). 

\item[Education Trust (1999).] {\it Dispelling the Myth: High 
Poverty Schools Exceeding Expectations}. Washington DC: Education Trust. 

\item[Eisenstein, Elizabeth L. (1983).] {\it The Printing Press 
as an Agent of Change} (Vols I-II). New York, NY: Cambridge University Press.

\item[Eisenstein, Elizabeth L. (2005).] {\it The Printing Revolution 
in Early Modern Europe}. New York, NY: Cambridge University Press. 

\item[Ericsson, K. Anders (2004).] {\it Deliberate practice and the 
acquisition and maintenance of expert performance in 
medicine and related domains}. Academic Medicine, 79.10, S70-S81.

\item[Fields, R. Douglas (2010).] {\it Change in the brains white 
matter}. Science 330.6005.768–769. Retrieved from 
http://dx.doi.org/10.1126/science.1199139.

\item[Fullan, Michael (1991).] {\it The New Meaning of Educational 
Change}. New York, NY. Teachers College Press.

\item[Fullan, Michael (2007).] {\it The New Meaning of Educational 
Change Fourth Edition}. New York, NY: Teachers College Press.

\item[Gaddis, John Lewis (2002).] {\it The Landscape of History, 
How Historians Map the Past}. New York, NY: Oxford University Press.

\item[Gladwell, Malcolm (2008).] Outliers, The Story of Success. 
New York, NY: Little, Brown and Co.

\item[Glazek, Stanislaw D. (2008).] {\it Heuristic model of teaching}. 
Retreived from arXiv:0804.4796v2.

\item[Glazek, Stanislaw D. (2011).] {\it Edukacja XXI}. PAUza 108, 1-2. 
Retrieved from 

http://pauza.krakow.pl/108$_-$12$_-$2011.pdf. 
English translation retrieved from 

http://www.fuw.edu.pl/~stglazek/PAUza108$_-$12$_-$2011English.pdf. 

\item[Glazek, Stanislaw D. and Seymour B. Sarason (2006).] 
{\it Productive Learning}. Thousand Oaks, CA: Corwin Press.

\item[Grissmer, David and Ann Flanagan (1998).] {\it Exploring 
rapid Achievement Gains in North Carolina and Texas}. 
Lessons From the States. Denver: Educational Commission of the States.

\item[Hark, John Max (1893).] {\it John Amos Comenius, Exercises 
in commemoration of the three hundredth anniversary 
of his birth, 1592-1892. His private life and personal 
characteristics}. National Education Association, 
Journal of Proceedings and Addresses. Session of the 
Year 1892 held at Saratoga Springs, New York, 1893, 703-712. 
Retrieved from 
http://archive.org/stream/addressesproce1892natiuoft\#page/848/mode/2up.

\item[Hesselbein, F. (2011).] {\it My Life in Leadership}. 
San Francisco, CA: Jossey-Bass.

\item[Hughes, Thomas P. (1983).] {\it Networks of Power, 
Electrification in Western Society, 1880-1930}. 
Baltimore, MD: Johns Hopkins University Press.

\item[Janowski, A (2010).] {\it By\'c dzielnym i umie\'c si\c e r\'o\.zni\'c: 
Szkice o Aleksandrze Kami\'nskim}. Niezale\.zne Wydawnictwo Harcerskie, Warszawa. 

\item[Johnson, David W., Roger Johnson, and Edythe J. Holubec (1994).] 
{\it Cooperative Learning in the 

Classroom}. Alexandria, VA: 
Association for Supervision and Curriculum.

\item[Johnson, David W. and Roger Johnson (2014).] University of 
Minnesota. Research Works. Retrieved from 
http://www.cehd.umn.edu/research/highlights/coop-learning/. 

\item[Kami\'nski, Aleksander (2013).] {\it My\'sli o Polsce i Wychowaniu}. Muzeum 
Harcerstwa, Warszawa.

\item[Karwowski, W. (2005).] {\it Ergonomics and human factors: 
the paradigms for science, engineering, design, technology 
and management of human-compatible systems}. Ergonomics 48.5., 436-463.

\item[Landes, David S., Joel Mokyr and William J. Baumol (Eds.). 
(2012).] {\it The Invention of Enterprise: Entrepreneurship from 
Ancient Mesopotamia to Modern Times}. Princeton, NJ: (Princeton University Press.

\item[Lillrank, Paul (2010).] {\it Service Processes. Introduction to 
Service Engineering}. Gavriel Salvendy and Waldemar Karwowski (Eds.). 
Hoboken, NJ: John Wiley \& Sons.

\item[Lyons, Carol and Gay Su Pinnell (2001).] {\it Systems for Change in 
Literacy Education: A Guide to Professional Development}. 
Portsmouth, NH: Heinemann Press.

\item[Maciariello, Joseph A. and Karen E. Linkletter (2011).] 
{\it Drucker's Lost Art of Management: 

Peter Drucker's Timeless 
Vision for Building Effective Organizations}. MacGraw-Hill, New York.

\item[McDermott, Lillian and the Physics Education Group (1995).] 
{\it Physics by Inquiry: 

An Introduction to Physics and the 
Physical Sciences} (Vols 1 and 2). Hoboken, NJ: John Wiley \& Sons.

\item[National Commission on Excellence in Education (1983).] 
{\it A Nation at Risk: The Imperative for Educational Reform}. 
A Report to the Nation and the Secretary of Education, 
United States Department of Education April 1983. Retrieved 
from http://datacenter.spps.org/uploads/SOTW$_-$A$_-$Nation$_-$at$_-$Risk$_-$1983.pdf.

\item[National Research Council (2012).] {\it Discipline-Based Education 
Research: Understanding and Improving Learning in Undergraduate 
Science and Engineering}. Washington, DC: The National Academies Press.

\item[National Writing Project (2014).] Retrieved from http://www.nwp.org/.

\item[Newman, William M. and Walter G. Vincenti (2007).] {\it On an 
engineering use of engineering history}. Technology and 
Culture 48.1. pp 245-247. 

\item[Perrow, Charles (1999).] {\it Normal Accidents, Living with 
High-Risk Technologies}. Princeton, NJ: Princeton University Press.

\item[Petroski, Henry (1996).] {\it Invention by Design, How Engineers 
Get from Thought to Thing}. Cambridge, MA: Harvard University Press.

\item[Petroski, Henry (2006).] {\it Success through Failure, the paradox 
of design}. Princeton, NJ: Princeton University Press.

\item[Project Lead the Way (PLTW) (2014).] Retrieved from 
https://www.pltw.org/. 

\item[Reading Recovery Council of North America (2014).] 
Retrieved from www.readingrecovery.org. 

\item[Sheppard, Keith (2005).] {\it The history of the academic credit 
system in America and its impact on the development of 
science education}. Paper presented at International History 
and Philosophy of Science Teaching Conference July 15-18, 2005. Leeds University, UK. 

\item[Slavin, Robert E. (2011).] {\it Educational Psychology: 
Theory and Practice (Tenth Edition)}. Upper Saddle River, NJ: Pearson.

\item[Slavin, Robert E. (2014).] {\it Success for All Foundation}. 
Retrieved from http://www.successforall.org/. 

\item[Smith, George E. (2009).] {\it Testing Newtonian gravity then 
and now}. Suppes Lectures at Stanford University. Retrieved 
from http://www.stanford.edu/dept/cisst/visitors.html.

\item[Staley, David J. (2010).] {\it History and Future, Using Historical 
Thinking to Imagine the Future}. Lanham, MD: Lexington Books. 

\item[STHESCA (2011).] {\it Science, technology, higher education 
and society in the conceptual age}. Krakow, Poland. 
Retrieved from http://sthesca.eu/.

\item[Vincenti, Walter G. (1993).] {\it What Engineers Know and How 
They Know It: Analytical Studies from Aeronautical History}. 
Baltimore, MD. Johns Hopkins University Press.

\item[Wikipedia. (2014).] {\it Micha{\l} Kulesza (prawnik)}. Retrieved from
 
http://pl.wikipedia.org/wiki/Micha{\l}$_-$Kulesza$_-$(prawnik) 

\item[Wilson, Kenneth G. and Bennett Daviss (1994).] 
{\it Redesigning Education}. New York, NY: Teachers College Press.

\item[Wilson, Kenneth G. and Constance K. Barsky (1996).] 
{\it Applied research and development: support for continuing 
improvement in education}. Daedalus 127.4. pp. 233-258.

\item[Wilson, Kenneth G. and Constance K. Barsky
(1999).] {\it School transformation: a case for a missing role. 
The Challenge of School Transformation: What Works}. 
Nottingham, UK: University of Nottingham.

\end{description}

\end{document}